\documentclass[10pt,emulateapj,apj]{emulateapj}

\shorttitle{}
\shortauthors{Park, Gott, \& Choi}
\begin{document}
\title{Transformation of Morphology and Luminosity Classes of the SDSS Galaxies}
\author{ Changbom Park\altaffilmark{1}, J. Richard Gott III\altaffilmark{2}, 
\& Yun-Young Choi\altaffilmark{1}}
\altaffiltext{1}{Korea Institute for Advanced Study, Dongdaemun-gu, Seoul 130-722, Korea; cbp@kias.re.kr, yychoi@kias.re.kr}
\altaffiltext{2}{Department of Astrophysical Sciences, Peyton Hall, Princeton University, Princeton, NJ 08544-1001, USA}
\begin{abstract}
We present a unified picture on the evolution of galaxy 
luminosity and morphology.
Galaxy morphology is found to depend critically 
on the local environment set up by the nearest neighbor galaxy in 
addition to luminosity and the large scale density. 
When a galaxy is located farther than the virial radius from its 
closest neighbor, the probability for the galaxy to have an
early morphological type is an increasing function only of 
luminosity and the local density due to the nearest neighbor ($\rho_{\rm n}$).
The tide produced by the nearest neighbor is thought to be
responsible for the morphology transformation toward the early type
at these separations.
When the separation is less than the virial radius, 
i.e. when $\rho_{\rm n} > \rho_{\rm virial}$, its morphology depends 
also on the neighbor's morphology and the large-scale background density
over a few Mpc scales ($\rho_{20}$) in addition to luminosity 
and $\rho_{\rm n}$.  The early type probability keeps increasing 
as $\rho_{\rm n}$ increases if its neighbor is an early 
morphological type galaxy. But the probability decreases 
as $\rho_{\rm n}$ increases when the neighbor is a late type. 
The cold gas streaming from the late type neighbor 
can be the reason for the morphology transformation toward late type. 
The overall early-type fraction increases as $\rho_{20}$ increases 
when $\rho_{\rm n} > \rho_{\rm virial}$. This can be attributed to the hot 
halo gas of the neighbor which is confined by the 
pressure of the ambient medium held by the background mass.
We have also found that galaxy luminosity depends on $\rho_{\rm n}$,
and that the isolated bright galaxies are more likely to be
recent merger products.
We propose a scenario that a series of morphology and luminosity
transformation occur through distant interactions and mergers,
which results in the morphology--luminosity--local density relation.
\end{abstract}

\section{Introduction}
Diverse galaxy morphology is one of the cosmological mysteries 
in which the secret of structure formation and evolution might be hidden.
Galaxy morphology is known to be correlated with luminosity and spatial
environment, and there is a hope that knowledge on
the correlations among them may provide clues to unravel the mystery.
This line of study began in the 1930's.
Hubble \& Humason (1931) found that clusters were 
dominated by ellipticals and lenticulars and that environmental 
factors played an important 
role in determining the morphology of galaxies. Oemler (1974) found 
the morphology-radius relation; the late type galaxy fraction 
decreases with radius within a cluster. 
This relation was confirmed by Dressler (1980) who argued that 
the fraction of morphological types is a function of local galaxy density. 
Postman \& Geller (1984) extended this morphology-density 
relation down to the group environment. 
On the other hand, galaxy luminosity is also known to depend
on environment. An excess of bright galaxies in clusters has been pointed
out by Schechter (1976). It is now well-accepted that both amplitude
and shape of the luminosity function of the individual morphological
types vary with environment (Park et al. 2007).
Despite these empirically known relations, it is 
very complicated to disentangle independent 
factors from the interrelations among various physical parameters 
of galaxies and environment.

On the theoretical side, there have been many ideas about the mechanisms 
causing such observational effects.
Gunn \& Gott (1972) argued that S0 galaxies in 
great relaxed clusters like Coma were the result of spirals being 
stripped of gas by ram pressure stripping due to hot intra-cluster gas. 
High speed encounters of galaxies with other halos typically in clusters
cause impulsive heatings, called harassment (Moore et al. 1996),
and can transform spirals to early types.
Strangulation (Balogh, Navarro \& Morris 2000) is another mechanism
that can also transform morphology through decline of the star formation rate
(SFR) due to shut-off of the newly accreted gas when a galaxy enters
a cluster or group environment and looses its hot gas reservoir.
There is also observational and theoretical evidence that the general tidal
force field in clusters can transform infalling spirals to early
types (Moss \& Whittle 2000; Gnedin 2003).
It is also plausible for galaxy morphology to be largely determined
by the initial conditions when galaxies were formed.
Gott \& Thuan (1976) argued that elliptical 
galaxies were produced by a larger initial density  which would have higher 
density at turn-around and where star formation would be completed before 
collapse. Such larger initial density  would be more likely in a region 
that would later turn into a high density environment. 

It is important to note that all the nurture processes proposed
so far are basically effective in group or cluster environments.
However, there seems to be morphology and luminosity-determining
processes which work at all local densities in a continuous way.
This is supported by the fact that there is no feature in 
the galaxy property versus local density relation
at any local density when morphology is fixed (Park et al. 2007).
But it was found that the fraction of late morphological type 
decreases sharply above
the critical luminosity of about $M_r = -21.3 + 5{\rm log}h$ in the morphology
versus luminosity relation (Choi et al. 2007, hereafter Paper I).

Recently, the Two Degree Field 
Galaxy Redshift Survey (2dFGRS; Colless et al. 2001) and the Sloan 
Digital Sky Survey (SDSS; York et al. 2000) have been
extensively used to accurately measure the environmental 
effects on various physical properties of galaxies 
(Goto et al. 2003; Balogh et al. 2004a,b; 
Blanton et al. 2005a; Croton et al. 2005; Tanaka 
et al. 2004; Weinmann et al. 2006; among many others). 
In this paper, we extend our work in companion papers which studied 
the relations among various physical properties of 
galaxies in the SDSS spectroscopic sample (Paper I), and 
the environmental dependence of the physical parameters (Park
et al. 2007, hereafter Paper II). 
In particular, we will further inspect the findings that galaxy
morphology varies with both the density measured on a few Mpc scales
and the distance to the nearest bright galaxy, and that the
dependence of the early-type fraction on the nearest
neighbor distance is strongest when the distance is about $200 h^{-1}$ kpc.
It will be shown that this separation corresponds to the virial
radius of the typical galaxies used in the study.
We will perform an extensive study of the dependence of galaxy
morphology on luminosity, large-scale density, the local density
due to the closest neighbor, and morphology of the neighbor.
Galaxy morphology is found to depend on the various local environment
parameters in a more complicated way.
Taking into account the fact that galaxy morphology is closely
related with luminosity, we extended our work to inspect the morphology 
transformation in conjunction with luminosity class
transformation of galaxies.
For this purpose, we use large volume-limited 
samples divided into accurate morphological subsets of early 
and late type galaxies in several luminosity bins.

\section{Observational Data Set}
\subsection{Sloan Digital Sky Survey Sample}
We use a large-scale structure sample, DR4plus (LSS-DR4plus), 
of the SDSS (York et al. 2000; Blanton et al. 2003a; 
Fukugita et al. 1996; Gunn et al. 1998, 2006; Hogg et al. 
2001; Ivezic et al. 2004; Lupton et al. 2001; Pier et al. 2003; 
Smith et al. 2002; Stoughton et al. 2002; Tucker et al. 2006) 
from the New York University Value-Added Galaxy Catalog 
(NYU-VAGC; Blanton et al. 2005b). 
This sample is a subset of the SDSS Data Release 5 
(Adelman-McCarthy et al. 2007). Our major sample of galaxies used 
here is the LSS-DR4plus sample referred to as ``void0,'' 
which includes the Main galaxies (Strauss et al. 2002) with 
the $r$-band apparent magnitudes in the range $14.5 < r_{\rm Pet} <17.6$ and
redshift in the range $0.001<z<0.5$.
For the approximately 6\% 
of targeted galaxies that lack a measured redshift because of 
fiber collisions, we assign the redshift of the nearest neighbor. 

The $r$-band absolute magnitude $^{0.1} M_r$ used in this study
is the AB magnitude converted from SDSS magnitudes.
To compute colors, we use extinction (Schlegel et al. 1998) and
K-corrected model magnitudes. 
The superscript 0.1 means the rest-frame magnitude
K-corrected to redshift of 0.1 (Blanton et al. 2003b).
All of our magnitudes and colors follow this convention,
and the superscript will be subsequently dropped.
We also drop the $+5{\rm log}h$ term in the absolute magnitude.
We use the luminosity evolution correction of $E(z) = 1.6(z-0.1)$
(Tegmark et al. 2004).  We adopt a flat $\Lambda$CDM cosmology
with $\Omega_{\Lambda} = 0.73$ and $\Omega_m = 0.27$.

Completeness of the SDSS is poor for bright galaxies with $r < 14.5$
because of spectroscopic selection criteria (which exclude 
objects with very large flux within the fiber aperture) and 
difficulties of automatically measuring photometric properties of very 
extended sources. 
As described in detail in section 2.1 of Paper I, we add the 
missing bright galaxies and thereby extend the magnitude 
range by using various existing redshift catalogs which 
we match to the SDSS data. 
In total, 5195 bright galaxies are added to the void0 samples
within our angular sample boundaries (see Fig. 1 of Paper II). 
Volume-limited samples derived from the resulting catalog 
have nearly constant comoving number density of galaxies 
in the radial direction for redshifts $z\geq 0.025$.
We treat our final sample as effectively having no bright limit at 
redshifts greater than $z = 0.025$. More details about 
this sample can be found in Paper I.

To study the effects of environment on galaxy properties it is
advantageous for the observational sample to have the lowest possible
surface-to-volume ratio, so that boundary effects are minimized
(see Park et al. [2005] regarding similar effects in a topology analysis). 
For this
reason we trim the DR4plus sample as shown in Figure 1 of Paper II.
The three stripes in the Southern
Galactic Cap observed by SDSS are not used because of their
narrow angular extent.  These cuts leave approximately $4464$
deg$^2$ in the survey region.  Within our sample boundaries, we
account for angular variation of the survey completeness by using 
the angular selection function defined in terms of spherical polygons (Hamilton 
\& Tegmark 2004), which takes into account the incompleteness 
due to mechanical spectrograph constraints, bad spectra, or bright
foreground stars.
The resulting useful area (with nonzero
selection function) within the analysis regions is 1.362 sr.  
In our study it is very important to use volume-limited samples 
of galaxies that maintains uniform sampling in space and luminosity.
The absolute magnitude and redshift limits of a set of our volume-limited
samples are shown in Figure 2 of Paper I.  
The definitions for the samples are summarized in Table 1.
We mainly use the $D4$ sample  which includes 74,688 galaxies with
$M_r \leq -19.5$.
The $D5$ sample, used in Figure 5, includes 80,479 galaxies with
$M_r \leq -20.0$.

\begin{deluxetable*}{lcccc}
\tabletypesize{\small}
\tablecaption{Volume-limited Samples}
\tablecolumns{6}
\tablewidth{0pt}
\tablehead{
\colhead{Name} &\colhead{Absolute Magnitude} &\colhead{Redshift}&
\colhead{Distance \tablenotemark{a}}&
\colhead{Galaxies}
}
\startdata
D3& $M_{\rm r}<-19.0$& $0.025<z<0.06869$& $74.6<R<203.0$& 49571\\
D4& $M_{\rm r}<-19.5$& $0.025<z<0.08588$& $74.6<R<252.9$& 74688\\
D5& $M_{\rm r}<-20.0$& $0.025<z<0.10713$& $74.6<R<314.0$& 80479\\
\enddata
\tablenotetext{a}{Comoving distance in units of $h^{-1}$Mpc}
\end{deluxetable*}

\subsection{Morphology Classification}

An important feature of this work is the use of large, accurate 
morphology subsets. 
We first classify morphological types of galaxies using the prescription of
Park \& Choi (2005). Galaxies are divided into early (ellipticals and 
lenticulars) and late (spirals and irregulars) morphological types 
based on their locations
in the $u-r$ color versus $g-i$ color gradient space and also in the 
$i$-band concentration index space. 
Our study is about the morphological transformation
between the early and late types, and not within the subclasses.
To measure the color gradient and the concentration index,
we use the $g$- and $i$-band atlas images and employ 
basic photometric parameters measured by the Princeton/NYU group
\footnote{\url{http://photo.astro.princeton.edu}}.
The boundaries between the two types 
in the three-dimensional parameter space 
is determined in such a way that the classification best reproduces 
the visual morphology classification.
The resulting morphological classification has completeness and 
reliability reaching 90\%, as claimed by Park \& Choi (2005). 
When photometry is excellent and galaxy images are well-resolved,
more information from surface brightness fluctuations can be added
for morphology classification. But this is certainly not the case
near the faint limit ($r=16.5\sim 17.6$) of the SDSS sample we use.

Our automatic classification scheme does not perform well
when an early type galaxy starts to overlap with other galaxy.
This is because the scheme excludes galaxies with very low concentration
from the early type class and blended images often erroneously give low
concentration.
Since we are investigating the effects of close pairs on galaxy
luminosity and morphology, this problem in the automatic classification 
has to be remedied.
In the case of the volume-limited sample D4 we perform an additional visual
check of the color images of galaxies to correct misclassifications 
by the automated scheme for about $20,000$ galaxies
located at the local density $\rho_{\rm n} /{\bar \rho} > 100$ 
(see below for the definition of $\rho_{\rm n}$).
In this procedure we changed the types of the rare galaxies which are blue
but elliptical galaxies, blended or merging ones, or dusty edge-on spirals.
Some non-sense objects are removed from the samples, and those with wrong
central positions are corrected.

\subsection{Local environment}

To find relations between intrinsic physical properties of galaxies
and their environment we require a well-defined and robust measure of
environment.  We seek to use environmental parameters that are
defined directly from observational data, and 
characterize the full range of galaxy environments, from the most 
massive clusters to voids. Most previous studies have used the local
galaxy number density as a measure of local environment.
In the present study we will consider three kinds of environment. One is
the mass density described by many neighboring galaxies over a few Mpc scale.
This is called the large-scale background density.
Another is the local mass density attributed to the closest neighbor
galaxy.
The third is the morphology of the closest neighbor galaxy.

The background density at a given location of a galaxy
is measured by 
\begin{equation}
\rho_{20}({\bf x})/{\bar\rho} = \sum_{i=1}^{20} \gamma_i L_i W_i(|{\bf x}_i -
{\bf x}|)/{\bar\rho},
\end{equation}
using the luminosity $L$ of the closest twenty galaxies in 
a volume-limited sample.
Here the mass associated with a galaxy is assumed to be proportional to the
luminosity of the galaxy. 
The mean mass density within a sample of the total volume $V$
is obtained by
\begin{equation}
{\bar\rho} = \sum_{\rm all} \gamma_i L_i /V,
\end{equation}
where the summation is over all galaxies in the sample.
The mean density of each volume-limited sample is nearly 
a constant when the magnitude limit of 
the sample is fainter than $M_r = -19.5$.
Only the relative mass--to--light ratios ($\gamma$) for early and 
late types are needed.
We assume $\gamma({\rm early}) = 2\gamma({\rm late})$
at the same $r$-band luminosity. This is our choice of the connection
between luminosity and morphology with the host halo mass.
It is based on the fact that the central velocity
dispersion of early type galaxies brighter than about $M_r=-19.5$ is about
$\sqrt{2}$ times that of late types (Fig. 4b of Paper I). It is also consistent 
with our finding that the RMS velocity difference 
between a galaxy and its neighbor galaxies is about 1.5 larger 
for the early types (see section 2.4).
We find the mean mass density
\begin{equation}
{\bar \rho} = 0.0223 \pm 0.0005 (\gamma L)_{-20},
\end{equation}
where $(\gamma L)_{-20}$
is the mass of a late-type galaxy with $M_r = -20$.
We use the spline-kernel weight $W(r)$ for the background density 
estimation as in Paper II.
We vary the size of the spline kernel, $h_s$, to include
a fixed number of galaxies within the kernel weighting.  
This adaptive smoothing kernel is often used in smoothed particle 
hydrodynamics simulations. We use the spline kernel because it is centrally
weighted, unlike the tophat or cylindrical kernel, and 
has a finite tail, unlike the Gaussian.  

Because this smoothing kernel is adaptive and centrally weighted, 
it has the important feature that it does not oversmooth dense 
regions, thereby blurring 
the `walls' into the `voids'.
Maps of the observed redshift space distribution of galaxies 
support this picture (see Fig. 3 of Paper II).
The adaptive nature of the smoothing kernel makes
the physical smoothing scale systematically
vary with environment, from a small smoothing scale in clusters to a
very large smoothing scale in voids.
Since the high density regions collapse while the under dense
regions expand as the universe evolves,
the adaptive kernel allows more uniform smoothing scale in the
`initial' conditions compared to the method adopting a fixed-scale
at the present epoch.

In our study the number of galaxies required
within the smoothing volume is set to $N_s =20$, which we find is 
close to the smallest number yielding good local density estimates.
When searching for the spline radius containing 
the required number of galaxies, 
the smoothing volume can hit the sample boundaries or
masked regions due to bright stars or galaxies.  In addition to this
trouble, the angular selection function varies across the sky even
within the survey region.  We first check if the location where the
background density is being measured is within the survey area, and then
ignore the survey geometry and angular selection function to search
for the nearest 20 galaxies in order to find the spline radius and
thus the local density using equation (1).  
We then randomly throw 1000 test particles
into the smoothing volume around the location, and calculate the sum of
normalized weights of the test particles assigned by the spline 
kernel, the angular selection function, and the radial boundaries.
The final local density is given by the first estimate divided by the
sum of weights. 

We have estimated the accuracy of the $\rho_{20}$ measure of the
background density. We identified dark halos with mass more than about 
$1\times 10^{12} h^{-1} M_{\odot}$ (Kim \& Park 2006) from a large $\Lambda$CDM 
N-body simulation using $2048^3$ particles made by Park et al. (2005),
and measured the dark halo number density $\rho_{20}$ in real and redshift spaces.
The error in the redshift space $\rho_{20}$ is found to be
12 and 31\% at densities $\rho_{20}/{\bar\rho}= 0.5$ and 10, respectively.

In Paper II, only the $L_\ast$ galaxies with $M_r = -20 \sim -21$ 
were used to trace the number density field.
The background density $\rho_{20}$ here differs from that
of Paper II in that it represents the mass density rather than 
the number density and that the density tracers include all 
galaxies brighter than the faint magnitude limit of a sample.
As a result, $\rho_{20}$ distinguishes high density regions 
from low density regions much better. 
When the $D4$ sample is analyzed, the galaxies tracing the mass density
are all galaxies brighter than $M_r = -19.5$. 
The median radius of the smoothing kernel is $h_S = 2.96 h^{-1}$ Mpc
and its 68\% limits are $h_S = 1.96$ and $4.38 h^{-1}$ Mpc.

The small-scale density experienced by a target galaxy located 
at ${\bf x}$ is estimated by
\begin{equation}
\rho_{\rm n}({\bf x})/{\bar\rho} = 
3\gamma_{\rm n} L_{\rm n} /4\pi r_p^3 {\bar\rho},
\end{equation}
where $r_p$ is the projected separation of the nearest neighbor galaxy 
from the target galaxy. The density due to the nearest neighbor used
in our work does not represent the small-scale galaxy number density,
but rather the local mass density given by the nearest neighbor itself.
The method to find the nearest neighbor is described
in the next section. Besides the nearest neighbor in terms of separation, 
we also adopted the neighbor galaxy giving the maximum 
value of $\rho_{\rm n}$ or the galaxy
causing the maximum tidal energy deposit per unit mass 
(Binney \& Tremaine 1987). We use a formula for the tidal energy deposit
per unit mass 
\begin{equation}
\Delta e \propto M_{\rm n}^2 a^2 /\Delta v_{\rm n}^2 (r_p^2 + a^2)^2,
\end{equation}
where $M_{\rm n}$ is the mass of the neighbor, $\Delta v_{\rm n}$ is the velocity
difference, and the size (the RMS radius) of the target galaxy $a$ is set to 10 times
the Petrosian radius. This combines the formulae for distant and 
penetrating encounters.
When we used these criteria to choose the neighbor, 
our results were essentially the same.
This is because the closest neighbors still dominate the latter two cases.
The fraction of neighbors selected differently in these three
nearest neighbor selection schemes is only about 15\%. 
We therefore present our results using only $\rho_{\rm n}$ of the nearest neighbor.

About 6\% of galaxies in our samples missed the spectroscopic fibers
due to the minimum fiber separation of 55$''$ per tiling in the SDSS.
When we estimated $\rho_{20}$ or $\rho_{\rm n}$, we included those galaxies with 
borrowed redshifts from the nearest neighbor on the sky.
To estimate the errors in $\rho_{20}$ and $\rho_{\rm n}$ due to this treatment
we artificially increased the fraction of galaxies with 
borrowed redshifts in our sample and compared the results with the original
density estimates.
We first selected the galaxies who have measured 
redshifts and have neighbors within 55$''$.
The redshift of the nearest neighbor which has a measured redshift is assigned 
to these galaxies, neglecting their own redshifts
so that a total of 6\% of all galaxies in the sample D4 received new
redshifts by this procedure.
We found the ratio of $\rho_{20}$ calculated from the degraded sample
to the original density had $1.00\pm 0.12, 1.01\pm 0.15$, and $1.04\pm 0.21$ 
when $\rho_{20}/{\bar \rho} \le 3, 3\sim 20$, and $\ge 20$, respectively. 
This tells that assigning the nearest neighbor's redshift makes 
$\rho_{20}$ slightly overestimated in high density regions, 
and the scatter is between $10\sim 20$\%. 
This error is less than that due to the redshift space distortion. 
We also checked the fraction of the nearest neighbors that were selected 
differently in the degraded sample. 
It is found that 15 and 7\% of the nearest neighbors
are selected differently for the pairs with separations less 
than 0.1 and 1 $h^{-1}$Mpc.

\subsection{The nearest neighbor}
For a given target galaxy with absolute magnitude $M_r$ selected from a 
volume-limited sample its nearest neighbor is found in the following way.
The closest neighbor is the galaxy brighter than $M_r+0.5$ with the smallest 
projected
separation across the line of sight from the target galaxy 
and with a radial velocity difference less than $V_{\rm max}$.
In the case of D4 containing galaxies brighter than $M_r=-19.5$, 
we study only those target galaxies brighter than $M_r=-20.0$ 
so that their neighbors are complete.
These criteria are empirically found from the sample.
When we adopt galaxies fainter more than 0.5 magnitude as neighbors, our
conclusions remain the same but our statistics become worse 
as the number of target galaxies becomes smaller.
It also reflects the fact that the effects of the neighbor are more
prominent when the neighbor is brighter than the target galaxy.

To determine $V_{\rm max}$ we searched for all neighbor 
galaxies with velocity difference less than 1000 km s$^{-1}$ 
with respect to each target galaxy and with magnitude 
not fainter more than 0.5.
Figure 1 shows the velocity difference between the target galaxies
with $M_r = -19.5 \sim -22.0$ and neighbor galaxies (in the D3 sample)
brighter than $M_r =-19.0$ as a function of projected separation.
\begin{figure} 
\center
\includegraphics[scale=0.4]{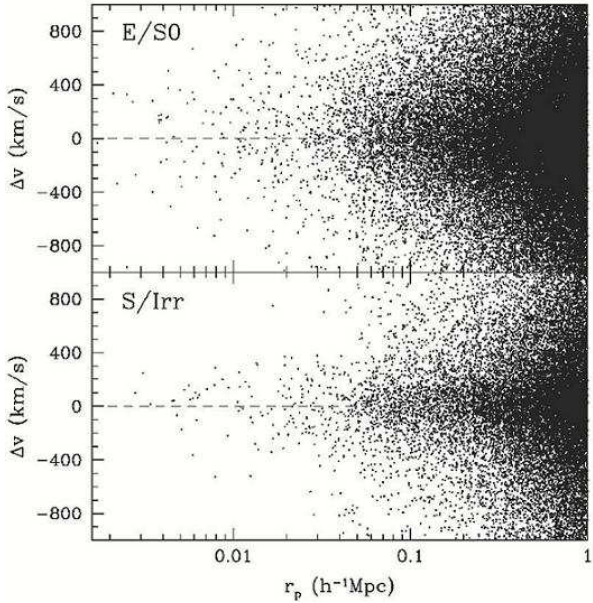}
\caption{Velocity difference between the target galaxies with $-19.5>M_r>-22.0$
and their neighbors brighter than $M_r +0.5$ as a function of the projected
separation. Most close neighbors
have velocity difference less than 600 and 400 km s$^{-1}$ when
the target galaxy is an early (E/S0) and 
late (S/Irr) morphological type, respectively.
The bottom panel for the late type galaxy indicates that contamination
due the interlopers are serious when $\Delta v > 600$ km s$^{-1}$.}
\end{figure}
Only the galaxies with measured redshifts are used.
It was found that the RMS velocity difference of the neighbors 
is nearly constant out to 
the projected separation of $50 h^{-1}$ kpc, and is 255 and 169 km s$^{-1}$
for early and late type target galaxies, respectively.
We adopt $V_{\rm max}=600$ and $400$ km s$^{-1}$ for the early
and late type target galaxies. 
These limits correspond to about 2.3 times the RMS values.

\section{Results}

\subsection{Morphology Transformation}
We first choose galaxies in the D4 sample with absolute magnitudes 
from $M_r=-20.0$ to $-20.5$.
They are basically $L_\ast$ galaxies (see Table 2 of Paper I
for the Schechter luminosity function parameters of the SDSS galaxies).
The large-scale background environment is divided into  high 
($\rho_{20}/{\bar\rho}\ge 20$), medium ($3\sim 20$), and low ($\le 3$) cases.
We then divide the small-scale neighbor environment
into two cases when the closest neighbor is of early or late morphological 
type. 
Then we look at the morphology of the target galaxy as a function 
of the neighbor density $\rho_{\rm n}$ when its luminosity, background density, 
and neighbor morphology are all fixed.
We first show Figure 2 for an intuitive understanding of the effects of
the neighbor's morphology and distance on the morphology of
the target galaxy. At a fixed luminosity bin the probability of a galaxy
to be an early type is a sensitive function of the projected separation
of the neighbor.
It also depends critically on the neighbor's morphology at separations
shorter than about $500 h^{-1}$ kpc.
However, at larger separations the dependence of $f_E$ on the neighbor
morphology disappears.

\begin{figure} 
\center
\includegraphics[scale=0.4]{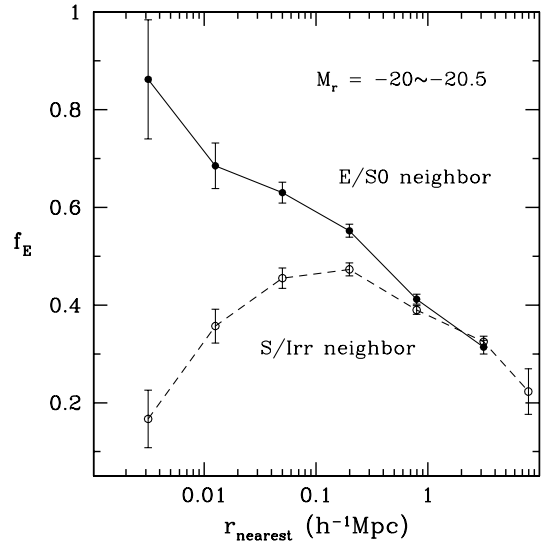}
\caption{Fraction of early (E/S0) morphological type galaxies as a function of 
the distance to the nearest neighbor. The luminosity of the target galaxies
is fixed to  $M_r =-20.0\sim -20.5$. The solid line with filled dots
is the case when the nearest neighbors are early types, 
and the dashed line with open circles is for the late (S/Irr) type 
neighbor cases.
}
\end{figure}

\begin{figure*}
\center
\includegraphics[scale=0.6]{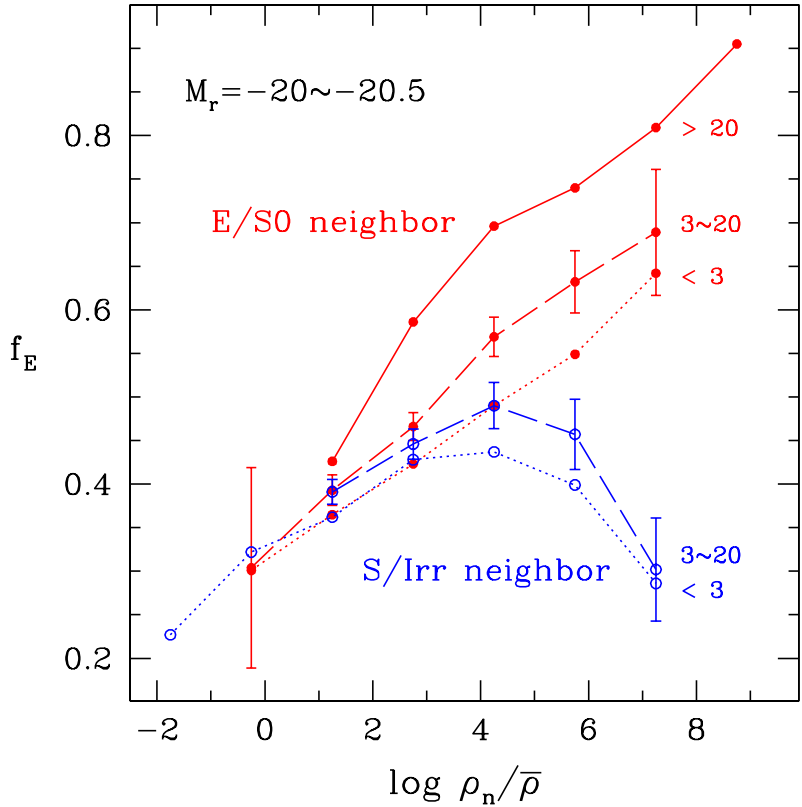}
\caption{Fraction of early (E/S0) type galaxies as a function of the small-scale
density due to the nearest neighbor. The luminosity of the target galaxies
is fixed to $M_r =-20.0\sim -20.5$. Each curve represents the case of a fixed
large-scale background density estimated by using 20 nearest galaxies
brighter than $M_r=-19.5$. The range in $\rho_{20}/{\bar\rho}$ is labeled
at the right of each curve. The upper three curves are the cases when
the nearest neighbor is an early type galaxy, and the lower two cases are
when the nearest neighbor is a late (S/Irr) type.
}
\end{figure*}

To obtain a deeper physical understanding we now use the mass density due
to the neighbor relative to the mean density of the universe
as the abscissa of Figure 3.
On the right side of the figure the upper three curves are the cases when
the nearest neighbor is an early type and the lower two curves are for
late type neighbor cases. The numbers next to the curves are the ranges
of $\rho_{20}/\bar{\rho}$.
It can be immediately noted that
the galaxy morphology is still a very sensitive function of the environment determined
by the nearest neighbor even if both luminosity and the large-scale density environment 
are fixed. According to the morphology--`density' relation (cf. Fig. 5 of Park et al. 2007)
galaxies are mostly of late type at very low (large-scale) densities and of early type
at very high densities. However, Figure 3 tells that the majority of galaxies are
of early type even at low densities if they have close early type neighbors, and that
the majority of galaxies are of late type even at high densities if they are very
isolated from other galaxies of comparable or brighter luminosity.
Figure 3 tells us more facts.
When $\rho_{\rm n}/{\bar\rho}\le 10^2$, the probability
$f_E$ for a galaxy to be an early type does not depend on 
the neighbor's morphology.
Its dependence on $\rho_{20}$ is also weak (but discernable). It is only the
neighbor density $\rho_{\rm n}$ (besides luminosity) which significantly
affects $f_E$;
the probability monotonically increases as $\rho_{\rm n}$ increases.
Hydrodynamic or radiative effects of the neighbor are not likely to be
responsible for this increase because there is no neighbor 
morphology dependence.
The effects acting in the environments like cluster or group can not be 
responsible for the increase of early type fraction
because the dependence on the large-scale background is insignificant
in this range of $\rho_{\rm n}$.
A possible mechanism responsible for this is the tidal effects by
the neighbor because the net effects of tide depend
only on kinematic quantities
such as separation, mass, size, and relative velocity (eq. 5). 
Tidal effects can accelerate the consumption of the cold gas in
galaxies and tend to transform late types into early types.

It will be interesting to quantitatively estimate the strength of the tidal effects 
generated by a neighbor located at about the virial radius.
Consider a galaxy having mass $M_0$ within radius $R_0$. Its gravitational
binding energy is $E_b \approx G M_0^2 /R_0$. 
The tidal energy deposit by a neighbor galaxy with mass $M_{\rm n}$ and relative
velocity $\Delta v_{\rm n}$ is given by equation (5) times $M_0$. When the separation
between the galaxies is much larger than the RMS radius of the galaxy,
the tidal energy deposit relative to the binding energy is given by
\begin{equation}
\Delta E /E_b \approx GM_{\rm n}^2 a^2 R_0 /M_0 \Delta v_{\rm n}^2 r_p^4.
\end{equation}
If we adopt $M_0 = M_{\rm n} = 1\times 10^{12} h^{-1} M_\odot, 
r_p = R_0 = 2a = 300 h^{-1}$ kpc, 
and $\Delta v_{\rm n}=$ 100 km/s, we obtain $\Delta E/E_b = 0.36$. Therefore,
the tidal effects between the dark halos of equal mass galaxies
is not negligible at the separation of about the virial radius (see below).
But the tidal force of the neighboring galaxy probably do little directly
to the stellar and gas components of the galaxy because of small 
$a$ and $R_0$ for these components.
It remains to examine if the rearrangement of the dark matter in the halo 
during the distant encounter can perturb the embedded late type galaxy 
significantly enough to accelerate the consumption of its cold gas.

When $\rho_{\rm n}/{\bar\rho}\ge 10^3$, however,
the probability at a fixed luminosity 
depends on all parameters, $\rho_{\rm n}, \rho_{20}$, and
neighbor's morphology, in a complicated but understandable way.
When galaxies have a late type closest neighbor, their 
$f_E(\rho_{\rm n})$ deviates
from that of the galaxies with early type neighbors at 
$\rho_{\rm n}/{\bar\rho} \approx 10^3$, and starts to drop at 
$\rho_{\rm n}/{\bar\rho} \geq 10^4$.

It is very important to note that this transition happens at the
virial density. 
The virialized density is $\rho_{\rm virial}/{\bar\rho}=
18\pi^2 /\Omega_m (H_0 t_0)^2= 766$ relative to the mean
density in the model universe we adopted (Gott \& Rees 1975).
Then equations (3) and (4) give the radius of the virialized region
$r_p \approx 240$ and $300 h^{-1}$ kpc for the late and 
early type galaxies with $M_r = -20$, respectively. 
The neighbor galaxy is holding both cold and hot gases within the virial
radius, and they can affect the fate of the galaxy falling into
the region.
If the neighbor is a late type galaxy, it is possible for its cold
gas to flow into a trapped early type galaxy and transform it to a late
type. 
If this picture is true, then one would expect to find
galaxies which appear to indicate
hydrodynamic and/or radiative interactions between close pairs of
galaxies before or without merger.
Figure 4 shows a few examples.
A $20 h^{-1}$ kpc scale bar at the rest frame is given to each panel.
The late type galaxies in the panels from (a) to (e) seem to be
transferring cold gas to their early type companions.

In panel (a) a long bridge extending from one spiral arm of a late
type galaxy ($z=0.0395$) is connected to an early type galaxy
($z=0.0397$). Despite its morphology the early type galaxy is very
blue ($u-r=1.73$), and has extremely strong emission lines indicating
active star formation.
The shape of this system has a surprising resemblance with a simulation
made by Cullen et al. (2007). They presented the observed spatial and
kinematic distribution of the atomic (HI) and molecular (CO)
gases in the spiral-elliptical interacting pair, Arp 104.
They also made a series of N-body simulations of the encounter between
a spiral and an elliptical galaxy to reproduce a system such as Arp 104.
Figure 8 of their paper is a best fitting case.
However, we note that the mass distribution in their simulation result
looks even closer to the system in panel (a).
An important lesson from their simulation is that a fraction of mass
in the disc of the late-type galaxy can be transferred to the early-type
component, and the spiral galaxy in panel (a) is probably feeding its
disk material to its elliptical neighbor, which can be responsible for
its spectral characteristics.

The panel (b) might be showing an example of the early-to-late transformation.
\begin{figure*}
\includegraphics[scale=0.9]{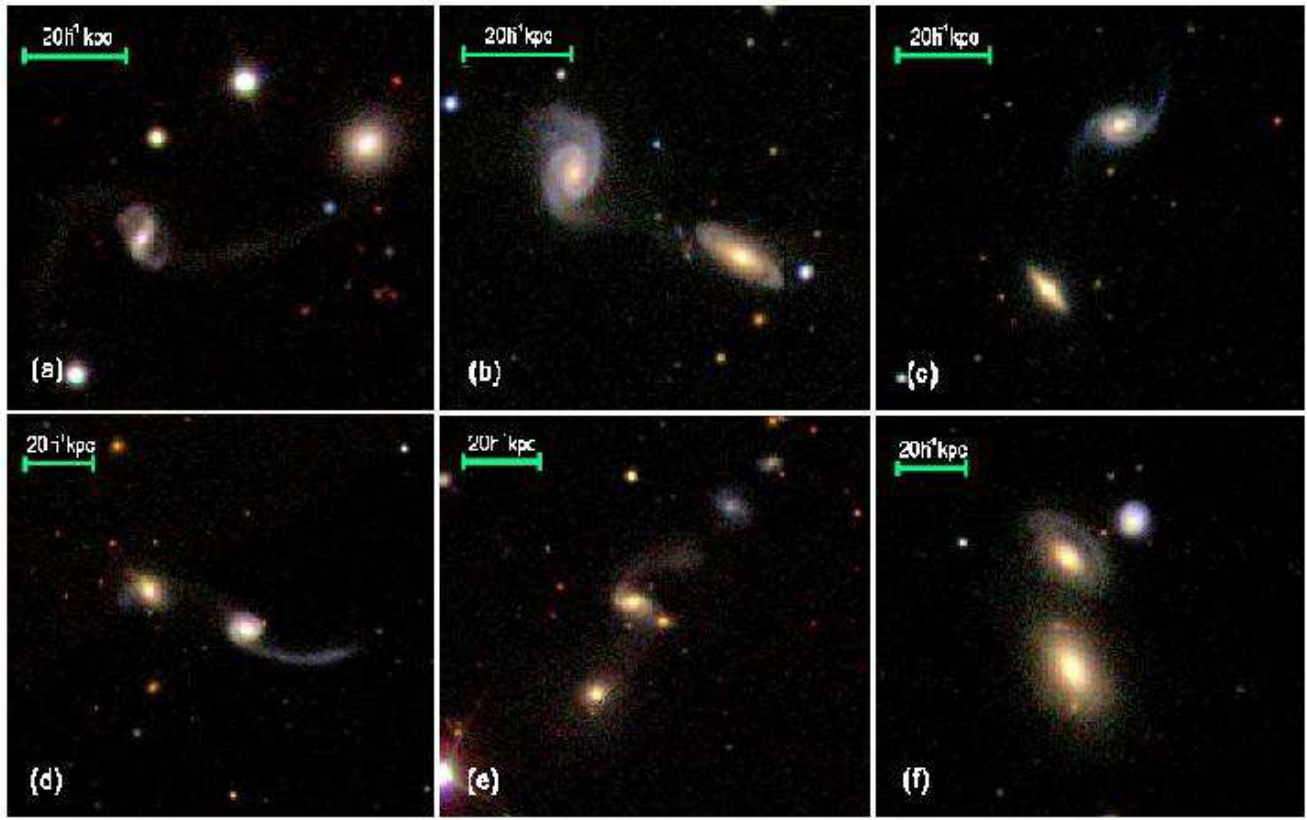}
\caption{({\it a}) - ({\it e}): Examples of interacting galaxies suggesting gas
transfer from a late-type galaxy to an early-type galaxy without or 
before merger. 
The red galaxy with two blue spiral arms, VCC1748, in panel (b)
might be a case showing the morphology transformation from early
to late type. 
({\it f}): An example suggesting that a bright early-type
galaxy affect the cold gas of a neighboring galaxy. The blue arc of the middle galaxy
is not seen at the side facing the bright elliptical galaxy below.
The position (RA, DEC) of the late type galaxy in each panel is 
(189.5657, 7.8204), (181.6186, 63.6299), 
(174.8445, 64.8037), (235.4750, 29.9913), (128.5264, 46.4387),
and (166.6741, 42.8220), from panel (a) to (f), respectively.
}
\end{figure*}
The galaxy VCC1748 (Binggeli, Sandage, \& Tammann 1985) with the recession
velocity of 11308 km s$^{-1}$ is at the lower right from the center.
It seems to be originally a red elliptical galaxy, but is now 
feathered with two blue spiral-looking arms.
Along the direction of its major axis is a blue spiral
galaxy VCC1752 having almost the same recession velocity of 11299 km s$^{-1}$.
The separation of two galaxies is only 34.5 $h^{-1}$ kpc, and the 
local density felt by VCC1748 due to VCC1752 is $3.5\times 10^5$ relative
to the mean density, much higher than the virial density.
One arm of the spiral galaxy is clearly connected to VCC1748 suggesting 
that the two arms are formed by the cold material flowed from VCC1752
along the bridge.
VCC1748 has previously been classified as SBa,
and seems to be an example that an elliptical galaxy 
transforms into a spiral when it closely encounters a gas-rich spiral
without or before a merger event.
The systems from panels (c) to (e) also indicate mass transfer
from a late type galaxy to an early type.

In fact, the mass transfer between galaxies during interactions 
has been widely studied both observationally and by numerical 
simulations.
Simulations indicate that the detailed interaction features and
the total mass transferred depend critically on the interaction
parameters (Toomre \& Toomre 1972; Sotnikova 1990; Mihos \& Hernquist 1994).
In particular, Walin \& Stuart (1992) pointed out that the transfer
of gas between galaxies can occur after they pass the point of
perigalacticon.
Observational evidence for mass transfer between galaxies has long been
suggested by many studies.
Knapp, Turner, \& Cunniffe (1985) examined the HI content of elliptical
galaxies and suggested that the gas has an external origin.
Counterrotating disks or bulges in spirals and ellipticals with dust
lanes can be interpreted as results of gas flows from the outside
(Bertola et al. 1998; Prada et al. 1996).
Many examples of ongoing mass transfer between interacting galaxies have
been studied. Some of them are Arp 105 (Duc et al. 1997), 
Arp 194 (Marziani et al. 2003), NGC 1409/1410 (Kell 2004), and
Arp 104 (Cullen et al. 2007).

The top three curves of Figure 3 indicate that when the nearest neighbor 
is an early type,
$f_E$ keeps increasing as $\rho_{\rm n}$ increases above the virial density.
In the light of the previous interpretation this would
be because the early type neighbor has a negligible amount
of cold gas and both its hot gas and tidal effects tend to
make the target galaxy falling within its virial radius an early type.
It is well-known that early type galaxies have extended X-ray halos
(Ellis \& O'sullivan 2006). 
The X-ray emission and the hydrodynamic effects of hot gas can only
accelerate consumption of cold gas in the close neighbor.
Since the bifurcation of $f_E$ for the late and early type neighbors 
occurs nearly at the virial density of the neighbor,
the hydrodynamic effects seem more responsible for the morphology
transformation than the radiative effects which should be less
sensitive to a particular scale.
This may not be true for galaxies much fainter than $M_r = -20$,
and the ionizing radiation from a bright elliptical galaxy could
affect the cold gas content of faint spirals beyond the virial
radius. This seems to happen at least to the satellites of bright
early and late type galaxies (Ahn, Park, \& Choi 2007).

The early type galaxies in panel (f) of Figure 4 support the argument
that the hot gas of an early type galaxy can remove cold gas from its
neighbor (in this case we have redshift only for the middle one.
We assume the two early type galaxies form a real interacting system.
They have nearly the same photometric redshifts).
The galaxy in the middle has a blue arc on the side facing
the spiral galaxy. 
However, the galaxy does not have such an arc on the opposite side
facing brighter elliptical galaxy.
We speculate that the hot gas of the elliptical neighbor at the bottom may be
responsible for non-existence of the second arc of the middle galaxy
as an elliptical neighbor can ionize or
push away the cold gas of its neighbor galaxy by a direct physical contact
and make it look like an early type (see Sofue 1994 for a demonstration
of such process in simulations).
We conclude that the physical properties of close neighbors are 
of paramount importance in the fate of a galaxy.

One can also learn from Figure 3 that $f_E$ is systematically higher
for higher $\rho_{20}$
when $\rho_{\rm n}$ is higher than the virial density.
The mechanisms suggested for morphology transformation to early type
due to large-scale environment
include the ram pressure stripping of a late type galaxy falling into a 
cluster holding hot gas (Gunn \& Gott 1972), removal of hot gas 
reservoir and stopping of cold gas supply (strangulation; Larson et al. 1980),
or impulsive heating due to numerous encounters with other galaxies in
high density environment (harassment; Moore et al. 1996).
Even though these mechanisms may be acting in the very high large-scale
density regions, 
none of these can be fully responsible for the large-scale background 
dependence we found because they all predict that $f_E$ is enhanced 
as $\rho_{20}$ increases independently of $\rho_{\rm n}$, which contradicts 
Figure 3.
(Note also that we cannot resolve the virialized region in the large-scale
density $\rho_{20}$.)
The fact that the strong $\rho_{20}$ dependence is observed only when
$\rho_{\rm n} > \rho_{\rm virial}$,
suggests that the effect of the background is indirect, and
the mechanism may be associated with a physical property of
the closest neighbor which in turn depends on the large-scale environment.
For example, this dependence can arise if the density and temperature
of the hot gas held by galaxies 
are higher in higher density environments.
Such a trend has been found only for the brightest group or cluster 
elliptical galaxies. 
They are reported to have higher X-ray luminosity at
given optical luminosity in higher density regions 
(O'sullivan et al. 2001).

Since it is very important to know the dependence of the X-ray emitting
hot gas of normal early type galaxies on the large-scale environment
to understand the dependence of $f_E$ on $\rho_{20}$, we
investigated the relation between X-ray luminosity of early types
and the large-scale background density.
We used X-ray data from the $\it ROSAT$ All-Sky Survey (Voges et al. 1999)
positionally matched with galaxies in the SDSS DR4plus catalog.
Among the matched sources there are 78 early types 
brighter than $M_r = -19.5$ within the volume of D4 sample,
we find that there exists a clear positive correlation between
the X-ray luminosity and $\rho_{20}$ at fixed optical ($r$-band) luminosity.
Our accurate density estimator enabled us to detect this correlation.
The dependence of $f_E$ on $\rho_{20}$ at
$\rho_{\rm n} > \rho_{\rm virial}$ can be explained 
by this correlation. 
The dependence is indirect and becomes effective only when
the hot gas of the neighbor galaxy starts to be influential
at separations between galaxies less than the virial radius.

So far, our study is done for the case when the luminosity of the target
galaxies is restricted to $M_r = -20.0 \sim -20.5$.
Figure 5 shows how the morphology transformation relation scales
with luminosity.
\begin{figure}
\center
\includegraphics[scale=0.4]{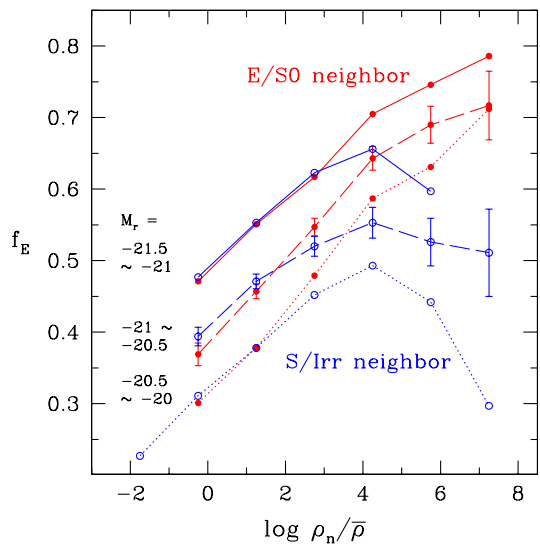}
\caption{Scaling of the morphology--neighbor density relation as a function
of target galaxy luminosity. Bottom two dotted curves used galaxies in D4 sample
with $M_r$ from $-20.0$ to $-20.5$, and having early (filled circles)
and late (open circles) neighbors. The upper four curves
used D5 sample.
}
\end{figure}
The bottom two (dotted) curves are the same as those in Figure 3 
but summed over all large-scale background densities.
The early-type fraction $f_E$ monotonically increases as $\rho_{\rm n}$
increases when the neighbor is an early type, but drops at
$\rho_{\rm n} /\bar{\rho} > 10^4$ when the neighbor is a late type.
The remaining curves are obtained from a deeper and brighter volume-limited
sample of galaxies with $M{_r}<-20.0$ (Sample D5 in Paper I).
The middle two curves are for galaxies with $-20.5 \geq M_r \geq -21.0$
having an early (higher $f_E$ at high $\rho_{\rm n}$) or late-type
closest neighbor.
The top two curves are for the half magnitude brighter galaxies
with $-21.0 \geq M_r \geq -21.5$.
We first note that the bifurcation of the early and late type neighbor
branches occurs at the same $\rho_{\rm n} /\bar{\rho} \approx 10^3$.
The neighbor's morphology is still an important environmental factor in
the morphology transformation at these brighter magnitudes.
Figure 5 tells that the whole morphology transformation relations 
as a function of $\rho_{\rm n}$ and neighbor's morphology, scale monotonically
with luminosity.
We have also confirmed this scaling at the fainter 
magnitude range from $M_r = -19.5$ to $-20.0$.
Above $M_r = -21.5$ a similar scaling is seen but the scaling
amplitude is suddenly very large. This is because the fraction of 
early type galaxies increases steeply above the magnitude of
$M_r \approx -21.3$ (See Fig. 8 of Paper I).
They are often the central galaxies in groups and clusters.

To summarize, a late type galaxy tends to become an early type due to
the tide exerted by its closest neighbor when it is separated by more than
the virial radius of the neighbor.
At separations less than the virial radius the galaxy still tends
to transform to an early type if the neighbor is an early type.
This trend is accentuated at higher large-scale background density.
But if the neighbor is a late type, the direction of morphology 
transformation tends to reverse and an early type can become a late type
as it approaches its bright spiral neighbor.
When one considers galaxies of different luminosity,
all behaviour remain the same and the probability for a galaxy to
be an early type simply scales with luminosity over the absolute magnitude
ranges we study.
Therefore, our understanding of galaxy morphology can not be complete unless
we understand how this luminosity--morphology relation originates.

\subsection{Luminosity Transformation}
From Figure 3
we were able to propose a number of physical processes playing 
roles in the galaxy morphology transformation.
However, one thing that Figure 3 can not tell us is what happens when
galaxies (located at very high $\rho_{\rm n}$) finally merge.
After a galaxy undergoes a major merger with its closest neighbor,
the next closest one will become the new closest neighbor.
The luminosity or mass of the merger product 
will be higher than those of the merging galaxies,
and the small-scale density $\rho_{\rm n}$
will drop greatly as the next closest neighbor as bright 
as the merger product will be usually far away.
If the galaxies had cold gas, the star formation must become 
relatively higher during the merger and the total cold gas content
in the merged one will be less than the sum of the cold gas
before the merger due to merger-induced starbursts.
The next major merger will occur with a more massive galaxy 
(as massive as the merged product) which tends to have
a lower fractional mass of cold gas.
As this approach-merge process repeats, the fractional mass of the merger
product in the form of cold gas will eventually vanish.

This scenario can explain why almost all most massive galaxies
are early types.
The scenario can be supported by our finding that galaxy luminosity depends
on $\rho_{\rm n}$ at a fixed large scale density.
When one inspects the effects of $\rho_{\rm n}$ or 
the nearest neighbor distance on various physical parameters of galaxies,
one finds only a very weak signal for all parameters 
except for luminosity (Y.-Y. Choi et al. 2007, in preparation).
Figure 6 shows the absolute magnitudes of galaxies in $D4$ as a function
of $\rho_{\rm n}$.
\begin{figure*}
\center
\includegraphics[scale=0.6]{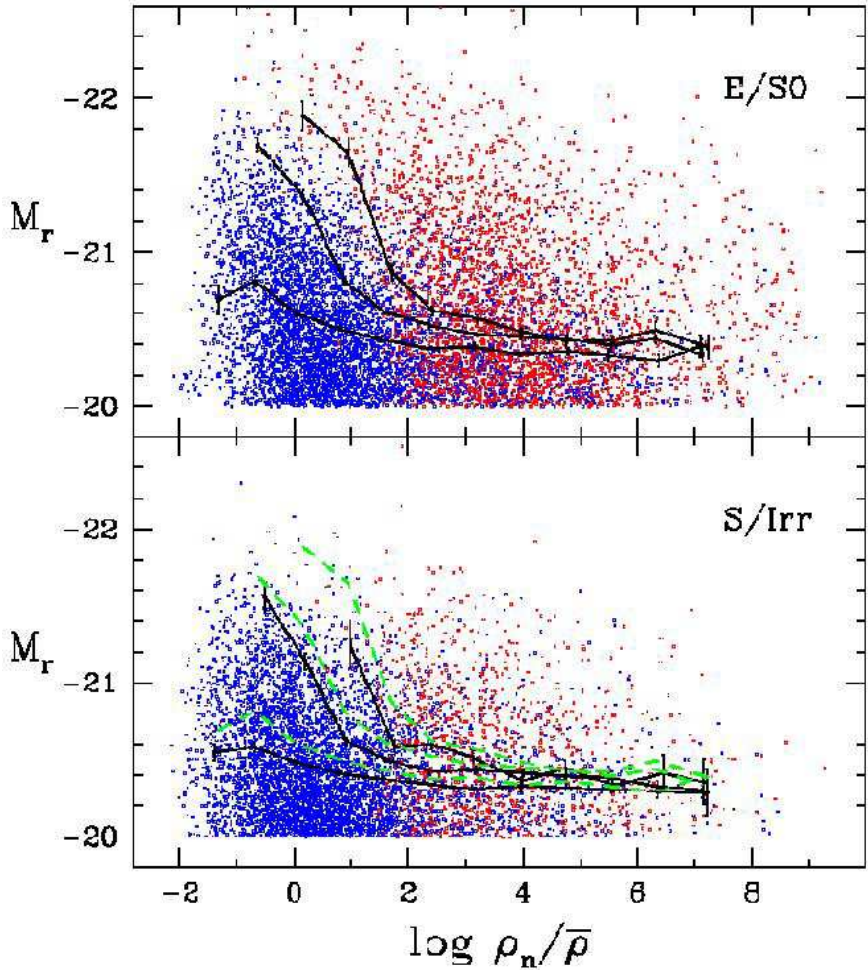}
\caption{The absolute magnitude of galaxies 
as a function of the neighbor density $\rho_{\rm n}$. 
The upper panel shows early (E/S0) type galaxies and the lower panel is
for late (S/Irr) types. Red dots are for galaxies located at 
$\rho_{20}/\bar{\rho} \geq 20$, and blue dots are for $<3$.
The top, middle, and bottom curves are the median magnitudes of galaxies 
located at
$\rho_{20}/\bar{\rho} \geq 20$, $3 \sim 20$, and $<3$, respectively.
Three curves for early types are repeated in the bottom panel
as dashed curves. 
}
\end{figure*}
The upper panel is for early type galaxies, and the bottom panel is for
late types.
As in Figure 3, we divide the large-scale background mass densities 
of galaxies into three bins according to their $\rho_{20}$.
We plot only those galaxies sitting at the highest $\rho_{20}$ (red dots)
or at the lowest $\rho_{20}$ (blue dots) bins to avoid overcrowding.
Three curves are the median magnitudes of galaxies in three $\rho_{20}$
bins as a function of $\rho_{\rm n}$, and
the curves for early types are repeated in the bottom panel
as dashed curves.

One can immediately notice that the luminosity of galaxies 
brighter than $M_r = -20$ and located in high $\rho_{20}$ regions 
is much higher at smaller $\rho_{\rm n} $. Namely, galaxies are
brighter as they are more isolated. And it 
is a strong function of $\rho_{\rm n}$ when $\rho_{\rm n} /\bar{\rho} < 10^2$.
But it can be also noticed that luminosity is nearly independent
of both $\rho_{\rm n}$ and $\rho_{20}$ when  $\rho_{\rm n} /\bar{\rho} > 10^3$,
and that all these behaviours are almost the same for both early
and late type galaxies 
(but early types are slightly brighter than the late types
in the magnitude range we explore).
We find the same behaviour of luminosity when fainter galaxies 
($M_r \leq -19.5$) are included.
To explain why isolated galaxies are brighter in high $\rho_{20}$ regions
and become faint as $\rho_{\rm n}$ increases, one can argue
that brighter galaxies preferentially form in higher large-scale
density regions and that galaxies become fainter by tidal stripping
as they approach each other.
However, this scenario is contradictory with the fact that 
the median luminosity is almost the same in different $\rho_{20}$ regions 
at $\rho_{\rm n} /\bar{\rho} > 10^3$,
and that galaxy luminosity
is independent of $\rho_{\rm n}$ at very high $\rho_{\rm n}$ where the tidal 
stripping should be more active.

An important fact that should be noted in Figure 6 is that
galaxies become more isolated (small $\rho_{\rm n}$) when 
$\rho_{20}$ is smaller or when luminosity is higher.
In high $\rho_{20}$ regions galaxies are on average much closer 
(higher $\rho_{\rm n}$) to one another than those in low $\rho_{20}$ regions.
Inevitably, the merge rate is relatively higher in higher $\rho_{20}$ regions.
We have visually identified merging (or close encounter) galaxies 
in the D2 sample brighter than $M_r = -20$. 
We looked for strongly interacting galaxies 
that show distorted bodies with connecting 
bridges and/or trails, and have magnitude difference less than 1.5. 
We included the cases that an early type galaxy 
is seen within the optical image of the other early type. We 
found, in the three density bins of $\rho_{20} <3$, $3 \sim 20$,
and $>20$, the fractions of galaxies undergoing major merger/interaction are
3.0, 4.8, and 6.8 \%, respectively.
This confirms the strong dependence of the major merger/interaction rate 
on $\rho_{20}$ (D. G. Kim et al. 2007, in preparation).
The galaxy merger rate in high density regions must be
higher than our estimate because there are more early type galaxies
at high densities and the merge between early types is more
difficult to notice as merging early types are less likely to show bridges or trails
than late types.
When a galaxy merges with another of equal mass, the merger
product will have twice the mass if the mass loss during the merger 
is negligible.
Since galaxies with twice the mass of the merging galaxies  will be rarer,
it will be more difficult for the merger product 
to find a neighbor that can significantly 
influence its morphology and luminosity.
Therefore, the merger product will jump to the upper (brighter)
left (low $\rho_{\rm n}$) part of each panel in Figure 6, as we argued before.
The rapid brightening of galaxies at $\rho_{\rm n} /\bar{\rho} \leq 10^2$
in the case of the highest $\rho_{20}$ bin, may be due to this process.
At what value of $\rho_{\rm n}$ will the galaxy find itself after merger?
It is reasonable to predict that the new $\rho_{\rm n}$ is typically less than
$\rho_{\rm virial}$.
If new $\rho_{\rm n} > \rho_{\rm virial}$, it implies two merging 
galaxies were already within 
a virial radius of a third one of about twice mass before merger.
This is not very likely because the more massive one in general impedes
their merger by tidal force and tends to swallow the little ones 
before they merge unless the pair were already ready to merge
before it fell into the more massive one.
This can explain why luminosity is nearly constant for galaxies at
$\rho_{\rm n} > \rho_{\rm virial}$.
Recently-merged galaxies usually do not fall at such $\rho_{\rm n}$.
Therefore, the  merger hypothesis can explain all features of Figure 5.

To find more dynamical evidence for this scenario we inspected the late
type galaxies showing indications of recent merger events.
Our prediction is that there are more galaxies showing post merger
features at $\rho_{\rm n} < \rho_{\rm virial}$ 
for a given luminosity and a fixed large
scale density.
For late type galaxies we define post merger features as large displacement
of the galaxy nucleus from the center, turmoil features, and/or very
close double cores.
The selection is subjective, but is made only for quite clear cases.
Only late types are used for higher reliability.
For galaxies with $M_r =-20.0 \sim -20.8$ and located in the
highest large-scale background density region ($\rho_{20}/\bar{\rho}>20$)
the fractions of galaxies showing the post merger features are
$3.3$ and $3.1$\% when $\rho_{\rm n} /{\bar{\rho}}
< 766$
and $766\le \rho_{\rm n} /{\bar{\rho}}<10^4$, respectively. 
For more luminous galaxies with $M_r = -20.8 \sim -21.6$
the fractions increase to 10.8 and 4.0 \%, respectively.
Therefore, there are more recently merged galaxies among isolated ones
than among those with close neighbors.

Figure 7 illustrates a merger in an isolated environment.
The $r$-band magnitude and redshift are given to
all galaxies brighter than $r=17.6$.
At the center of the picture is a spiral galaxy NGC3695 with two
cores indicating that it is at the ending stage of merger.
NGC3695 has $M_r = -21.27$, and is a blue spiral galaxy showing
very strong star formation activity. 
Nearly equal brightness of the two nuclei implies that two galaxies
of comparable mass have merged.
Since NGC3695 is completing merger, the next closest galaxy will become
a new neighbor.
The bright galaxy at the lower left of the picture is NGC3700.
Because NGC3700 has velocity difference of 12,300 km s$^{-1}$,
it does not qualify as a neighbor. 
The closest neighbor is found at the separation of
$1 h^{-1}$ Mpc, and NGC3695 is a good example of very isolated galaxy
which once was a close pair of galaxies.

To summarize, galaxies travel from the left to the right of Figure 6 at
a given luminosity.
During this travel, galaxies are affected by the morphology transformation
processes suggested by Figure 3, and can jump from one panel to the other.
When they reach very high $\rho_{\rm n}$, they undergo a merger and become brighter.
As they become brighter through mergers,
they will find themselves more isolated from their neighbors of
comparable mass.
As galaxies repeat such travel, the fractional mass in cold gas will
decrease and galaxies with higher mass will tend to be of early type.

\begin{figure*}
\center
\includegraphics[scale=0.7]{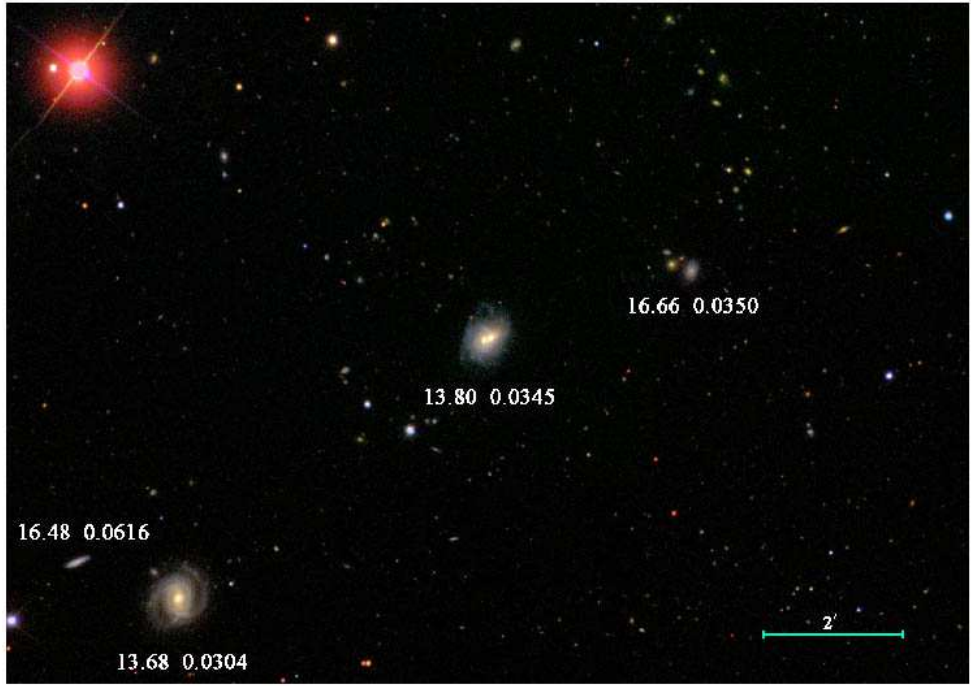}
\caption{An example of the isolated galaxy which once was a close
pair of galaxies. At the center is a spiral galaxy NGC3695 which
has two cores, indicating that it is a recent merger product.
Its closest neighbor is located at the separation of $1 h^{-1}$ Mpc
(not seen in this picture).
At the lower left is another spiral galaxy NGC 3700, which is a foreground
object not directly interacting with NGC3695.
The $r$-band magnitude and redshift are given to galaxies brighter than $r=17.6$.
}
\end{figure*}

\section{Discussion}
A full understanding of the origin of galaxy morphology will be obtained
if one finds the initial distribution of morphology and the rule of morphology
transformation.
The present study addresses the second issue.
If galaxies transform their morphology and luminosity in accordance with 
the unified scenario we propose, a number of galactic 
phenomena found in the past can be qualitatively explained.
Because the merger rate is higher in the higher large-scale density regions,
there are more bright merged galaxies in high density regions, which can explain
the luminosity-density relation (Paper II).
As the merger process repeats, galaxies become brighter but also tend to become
early types because the cold gas fraction decreases.
This can explain the luminosity--morphology relation.
Furthermore, the effect of hot gas of the neighbor on a galaxy 
orbiting within the virial
radius tends to make the galaxy an early type and its effects
are stronger in high large-scale density regions.
This can explain the morphology--density or morphology--(clustercentric) radius
relation (more discussion is given below).

In addition to these important relations our picture can explain
the following.

\begin{enumerate}
\item {\it Holmberg effect}
~~Holmberg (1958) discovered that the color of paired galaxies are closely
correlated. We predict that as two galaxies approach each other and 
enter into the virial radius, hydrodynamic and radiative interactions 
of cold and hot gases occurs. 
The primary galaxy tends to make the pair system have common properties.
The color of the secondary can become closer to that of the primary.

\item {\it Star formation rate--density correlation}
~~The fraction of star-forming galaxies is known to depend strongly on the
local density. Balogh et al. (2004a) showed that the fraction of $H\alpha$
emission-line galaxies is lower in regions
that are overdense at both 3.85 and $0.77 h^{-1}$ Mpc scales.
To a large extent this correlation is associated with the morphology density
relation. Our picture predicts that the cold gas content and thus
the SFR in galaxies will be lower in higher 
large-scale density
regions where the merger rate is higher and the early-type fraction is higher.
At the same time, the morphology of a galaxy is sensitive to that of 
its neighbor when it is located within the virial radius of the neighbor.
The SFR must be so too.
The neighbor's morphology depends on the background density
when the pair separation bin is small than the virial radius.
Therefore, the SFR should appear dependent on both small and large-scale
densities.
When the large-scale density and neighbor morphology are fixed,
the SFR of galaxies in close pairs is predicted to be depressed 
for the early type neighbor case or enhanced for late type neighbor case.

\item {\it Conformity in galaxy morphology type}
~~This is essentially the same as the previous two relations.
It has been observed both in group/cluster environment and at galaxy scales.
Weinmann et al. (2006) reported that a late-type central galaxy has a
significantly higher fraction of late type galaxies within its
halo than an early-type central galaxy of the same mass
(see also Hickson et al. 1984 and Ramella et al. 1987).
This can be just because of the strong hydrodynamic and radiative 
influences of the bright
central galaxy on the less bright surrounding galaxies which are orbiting
within the virial radius of the central one.
This galaxy morphology conformity is also observed within 
the halos of individual galaxies.
An elliptical galaxy is found to host a swarm of early type satellite
galaxies while a gas-rich spiral galaxy tends to hold many late type
satellites (H. Ahn et al. 2007 in preparation).
This is a scaled-down version of the phenomena observed in groups 
and clusters, and can be consistently explained by our picture.

\item {\it Morphology--density versus morphology--radius relation}
~~Galaxy morphology depends strongly on the clustercentric radius (Melnick \&
sargent 1977) or local density of galaxies in clusters (Dressler 1980).
It would seems impossible to find which of the factors is the more fundamental.
Our picture claims that it is the local density which really controls the
phenomena.
The large-scale background density determines the average distance to
the nearest neighbor and the merger rate which in turn determines the 
mean brightness and morphology of the neighbor.
The small-scale density due to the nearest neighbor determines the
strength of the tidal and hydrodynamic effects.
The virial radii of the group/cluster a galaxy belongs
to, or of its neighbor galaxy are statistically determined by local densities.
When a galaxy enters within the virial radius of the group/cluster or of 
the neighbor galaxy, it starts to interact with the hot/cold gas trapped 
within the radius.
Therefore, the virial radius has a fundamental importance in the evolution
of galaxy morphology.

\item {\it Morphology-dependent large-scale clustering}
~~It is well-known that the spatial distribution of the early-type galaxies has
higher clustering amplitude than the late types.
This can be because the early types preferentially form in high
density regions and their clustering is more biased than the late types
relative to the mass. However, our results imply that even if the initial
morphology of the newly-born galaxies is of late type, the luminous early
type galaxies are formed preferentially in high density regions
because the binary interaction is more frequent, 
the merger rate is higher, and the cold gas consumption is 
faster there.
Once the early type galaxies with hot gas appear in high density regions,
they will promote the transformation of morphology of neighboring spirals
through galaxy-galaxy interaction. 
Since high density regions are compact, this will result 
in the morphology-dependent clustering amplitude.
\end{enumerate}

Our picture on the transformation of galaxy morphology and luminosity
also explains some recent findings and can make predictions 
on galaxy properties.

\begin{enumerate}
\item {\it Existence of early types in underdense regions}
~~In our picture early type galaxies can form in low density regions
as well as in high density regions but at a lower efficiency through 
infrequent pair interactions and mergers.
It can be inferred from Figure 6 which shows there are much fewer galaxies with
$\rho_{\rm n}/{\bar\rho}>10^3$ in low density regions (blue points).
It predicts that, in underdense 
regions where there are more gas-rich late-type neighbors,
the very bright early types will be very rare and the 
fraction of blue star-forming early types will be higher.
This is exactly what is seen in Figure 11 and 12 of Paper II.
There are `normal' ellipticals even in very underdense regions
as demonstrated by the fact that the red-sequence of early types 
hardly changes as the large-scale density changes.
However, in underdense regions the red-sequence has a much shorter extension 
toward the bright magnitudes and the fraction of blue galaxies increases.
The fractions of early-type galaxies brighter than
$M_r =-21.5$ among those with $M_r < -19.5$ are 9.4, 6.6, and 2.8 \%
in the regions where $\rho_{20}/\bar{\rho} \geq 20$, $3\sim 20$, and $<3$,
respectively.
On the other hand,
the fractions of blue early type galaxies 
with $u-r<2.3$ and $M_r \leq -19.5$ are 1.2, 2.3, and 3.4 \% in those three
regions.
Properties of early type galaxies in low density regions are also studied by
Croton et al. (2005) and Rojas et al. (2005).
The simple scenario that postulates formation of all early type galaxies
in initially high density regions has a difficulty in explaining
existence of the normal and blue ellipticals in very low density regions.

\item {\it Isolated galaxies}
~~Isolated galaxies are often studied in the hope that the evolution of
galaxies can be understood with no contamination from environment-related 
processes.
Isolated galaxies are usually defined as those which do not have companions
brighter than a certain magnitude difference within a certain
projected distance and radial velocity difference.
They are considered to be passively evolving galaxies formed in isolation
via gravitational collapse of a primordial protogalactic cloud 
(Marcum, Aars, \& Fanelli 2004).
However, our scenario drawn from Figure 6 
predicts the most isolated galaxies are likely
to be those that experienced a major merger relatively recently
and whose internal evolution was greatly disturbed.
This is particularly so in high density regions or for very bright galaxies. 
The wish for freedom from environmental effects
will be approximately fulfilled only for the galaxies isolated at
very low densities, namely for the void galaxies.
For example, Marcum et al. (2004) have selected extremely
isolated elliptical galaxies from the Catalog of Isolated Galaxies 
(Karachentseva 1973) and found that four out of nine showed merger signatures.
It should also be noted that isolation of a galaxy is strongly correlated
with the large-scale density and characteristics of isolated systems
can be easily confused with those of underdense environment 
(i.e. Varela et al. 2004).

\item {\it Redshift evolution of morphology and luminosity}
~~Our scenario is consistent with the common wisdom that galaxies were on
average less massive in the past and that the late-type fraction
of galaxy morphology was higher at higher redshifts. It predicts that
galaxies located in high density regions should show relatively large differences 
in luminosity, color, SFR, and morphology as redshift changes
because the mean separation between galaxies is smaller and the merger
rate is higher there.
The fraction of early type galaxies should monotonically decrease and
the fraction of blue early type among early types should increase at
higher redshifts.
However, the energetic radiation from numerous active galaxies, which
are a minor population in our sample, might have critical roles in the
evolution at high redshifts.
Much less evolution will be observed for galaxies in low density regions
where the mean galaxy separation is large because the interaction between
galaxies is weaker and merger is less likely.
Correspondingly, the star formation--density relation will change
in such a way that the strongest SFR of galaxies be observed in higher
density regions as redshift increases (see Elbaz et al. 2007 for
an observational detection of this phenomenon).
Our scenario is also consistent with the recent finding from
a N-body simulation that internal
physical properties and spatial clustering of dark halos depend not only
on the mass but also on the assembly history (Gao, Springel, \& White 2005;
Wechsler et al. 2006; Gao \& White 2007; Croton, Gao, \& White 2007).
At high redshifts, galaxies at high densities are expected to be less 
depleted in cold gas and experience more frequent interactions
than those at low densities.
Since at higher redshifts the space was compressed 
but there were fewer number of massive
galaxies, the merger rate of objects above a given mass is expected
to evolve rather slowly (see Lotz et al. 2006 and Bell et al. 2006
for observational evidence)
\end{enumerate}

The galaxy morphology can also evolve due to the infalling gas in the
form of intergalactic medium or dwarf galaxies.
In less massive halo, the disk can be formed by cold flows ($\sim 10^{4-5}$K),
and in more massive halos the infalling gas is first heated by shocks to
near the virial temperature ($\sim 10^6$K) before it is accreted to form a disk
(Dekel \& Birnboim 2006; Keres et al. 2005). 
The relative importance between the galaxy-galaxy interaction
and the gas/dwarf accretion on morphology transformation will 
depend on the large-scale environment
and the time scales of gas cooling and interaction between galaxies. 
For examples, the importance
of the infalling gas will be relatively higher in low density regions where the
interaction rate is lower and there are more gas rich faint objects.
It is also possible for galaxies to transform their morphology between early and
late types through secular internal processes like exhaustion of cold gas and
aging of the stellar population. The time scale of such secular evolution
is probably longer than those of galaxy interaction or gas
accretion before the present epoch.
It will be interesting to compare the relative 
importance of all these processes in different environments.

\section{Conclusions}

We have analyzed a set of volume-limited samples drawn from the SDSS data
in order to understand the morphology--luminosity--environment relation.
Our previous study on the environmental dependence of galaxy
properties (Paper II) showed that most of the environmental dependence
of physical parameters disappears once luminosity and morphology
are fixed and that the morphological type of galaxies is affected by
both small-scale (due to the nearest neighbor) and large-scale densities.
In particular, a strong dependence of morphology on the nearest neighbor
distance was discovered when the distance was about $200 h^{-1}$ kpc. 
For galaxies with $M_r =-19.0$, which is the faint limit of the sample
D2 used in Paper II, the critical distance of $200 h^{-1}$kpc corresponds
to the neighbor density of $\rho_{\rm n} /{\bar \rho}=520$ or 1040 for
late or early type neighbor, respectively. This essentially
corresponds to the virial density. In the present work,
based on the lessons from the previous study, we divided the environment
into the large-scale background density, 
small-scale density due to the nearest neighbor,
and the nearest neighbor's morphology.
Our major findings are the following.

\begin{enumerate}
\item Transformation of galaxy luminosity and morphology classes is going
on in all large-scale density environments through interactions 
and mergers between close galaxy pairs.
\item At galaxy pair separations larger than the virial radius of the
neighbor galaxy, the morphological type of a galaxy is nearly independent 
of the neighbor's morphology. It is argued that
the tidal effects of the neighbor transform
late type galaxies into early types.
It is also hardly affected by the general tidal force field as indicated
by its insensitivity to the large-scale background density.
\item At separations from the closest neighbor shorter than the virial
radius, the probability of a galaxy to be an early morphological type
continues to rise as $\rho_{\rm n}$
increases if the neighbor is an early type.
The tidal, hydrodynamic, and radiative effects can result in this
phenomenon.
If the neighbor is a late type, however, the probability decreases 
as $\rho_{\rm n}$
increases and the galaxy tends to become a late type.
The tide and the hot gas may be competing with the cold gas flowing
from the neighbor in this circumstance.
\item The early type probability as a function of neighbor morphology,
large- and small-scale densities, 
scales systematically as the luminosity changes over
the magnitude range from $M_r = -19.0$ to $-21.5$.
The fraction increases sharply for $M_r \leq -21.5$.
\item It is found that galaxy luminosity depends sensitively
on the local mass density due to the nearest neighbor.
Isolated galaxies tend to be brighter and are more likely to be
recent merger products.
The merger-driven transformation of luminosity class can result
in these phenomena.
\item We present a unified scenario that the galaxy morphology
and luminosity classes change as galaxies approach one another 
and undergo a series of (mainly two-body) interactions and mergers.
As the average interaction and merger rates are a function of the
background density, the morphology--luminosity--density relation naturally
arises under this scenario.
\end{enumerate}

We were not able to study what happens to galaxy morphology and luminosity
when the large-scale background density $\rho_{20}$ 
exceeds the virial density.
This is because the smoothing scale used to calculate $\rho_{20}$,
which is $3.0^{+1.4}_{-1.0} h^{-1}$ Mpc, is 
larger than the size of the virial radius of clusters, which is typically
about $1 h^{-1}$ Mpc.
In a cluster environment, we expect to see morphology
transform of late type galaxies when they are 
within the virial radius of the cluster.

We emphasize that one needs to take into account factors other than
just one local density estimate when the effects of galaxy interactions
in different environments are studied.
In particular, the morphology of the closest neighbor is important
when the separation is smaller than the virial radius of the neighbor.
Without distinguishing the nearest neighbor by morphological type,
the study of the effects of interactions on color, SFR, or morphology
fraction will find almost a null signal because the early and late neighbors
will tend to induce opposite trends. We will further investigate such effects
of galaxy interactions in a forthcoming paper 
(Y.-Y. Choi et al. 2007 in preparation).

\acknowledgments
The authors thank H. B. Ahn, J. P. Ostriker, and M. S. Vogeley for
helpful comments.
CBP acknowledges the support of the Korea Science and Engineering
Foundation (KOSEF) through the Astrophysical Research Center for the
Structure and Evolution of the Cosmos (ARCSEC).
JRG is supported by NSF GRANT AST 04-06713.

Funding for the SDSS and SDSS-II has been provided by the Alfred P. Sloan 
Foundation, the Participating Institutions, the National Science 
Foundation, the U.S. Department of Energy, the National Aeronautics and 
Space Administration, the Japanese Monbukagakusho, the Max Planck 
Society, and the Higher Education Funding Council for England. 
The SDSS Web Site is http://www.sdss.org/.

The SDSS is managed by the Astrophysical Research Consortium for the 
Participating Institutions. The Participating Institutions are the 
American Museum of Natural History, Astrophysical Institute Potsdam, 
University of Basel, Cambridge University, Case Western Reserve University, 
University of Chicago, Drexel University, Fermilab, the Institute for 
Advanced Study, the Japan Participation Group, Johns Hopkins University, 
the Joint Institute for Nuclear Astrophysics, the Kavli Institute for 
Particle Astrophysics and Cosmology, the Korean Scientist Group, the 
Chinese Academy of Sciences (LAMOST), Los Alamos National Laboratory, 
the Max-Planck-Institute for Astronomy (MPIA), the Max-Planck-Institute 
for Astrophysics (MPA), New Mexico State University, Ohio State University, 
University of Pittsburgh, University of Portsmouth, Princeton University,
the United States Naval Observatory, and the University of Washington. 
{}
\end{document}